\documentclass[aps,pra,twocolumn,amsmath,amssymb,showpacs,nofootinbib]{revtex4-1}
\usepackage{graphicx}
\usepackage{amsmath}
\usepackage{bm}
\usepackage{xcolor}

\newcommand{\be}{\begin{equation}}  
\newcommand{\ee}{\end{equation}}
\newcommand{\ba}{\begin{array}}
\newcommand{\ea}{\end{array}}
\newcommand{\bea}{\begin{eqnarray}}
\newcommand{\eea}{\end{eqnarray}}

\newcommand{\bra}{\langle}
\newcommand{\ket}{\rangle}

\newcommand{\nn}{\nonumber}

\begin{document}

\title{Collective heat capacity for quantum thermometry and quantum engine enhancements\\
}

\author{C.L. Latune$^{1,2}$, I. Sinayskiy$^{1}$, F. Petruccione$^{1,2,3}$}
\affiliation{$^1$Quantum Research Group, School of Chemistry and Physics, University of KwaZulu-Natal, Durban, KwaZulu-Natal, 4001, South Africa\\
$^2$National Institute for Theoretical Physics (NITheP), KwaZulu-Natal, 4001, South Africa\\
$^3$School of Electrical Engineering, KAIST, Daejeon, 34141, Republic of Korea}

\date{\today}
\begin{abstract}
The performances of quantum thermometry in thermal equilibrium together with the output power of certain class of quantum engines share a common characteristic: both are determined by the heat capacity of the probe or working medium. After noticing that the heat capacity of spin ensembles can be significantly modified by collective coupling with a thermal bath, we build on the above observation to investigate the respective impact of such collective effect on quantum thermometry and quantum engines. We find that the precision of the temperature estimation is largely increased at high temperatures, reaching even the Heisenberg scaling -- inversely proportional to the number of spins. For Otto engines operating close to the Carnot efficiency, collective coupling always enhances the output power. Some tangible experimental platforms are suggested.   
\end{abstract}

\maketitle


Recent experiments realised proof-of-principle of some thermodynamic tasks based on single quantum systems, among which spins occupy a prominent place. 
 In particular, heat engines were realised using a single spin as working fluid \cite{Lindenfels_2019, Peterson_2018}. Temperature estimation of ultra cold gases via single quasispins of ceisum was performed in \cite{Bouton_2020}.  
  In \cite{Wang_2018, Liu_2019}, nano-thermometers were experimentally realised using spins of Nitrogen-vacancy centers (where the temperature change in the environment was mapped into magnetic signal through a magnetic nanoparticle).  
 %
Here we ask the following question: can collective spin effect enhance such thermodynamic tasks?

In quantum thermometry, most studies investigating collective effects rely on quantum phase transition \cite{Mehboudi_2015, Pasquale_2015, Salvatori_2014, Mehboudi_2016, Potts_2019}. Beyond that, some studies \cite{Mehboudi_2019, Guo_2015} investigate precision enhancement based on the small energy splittings emerging from interactions between subsystems contained in the probe. Here, we focus on a different collective effect: the collective coupling between a bath and an ensemble of spins. One study \cite{Gebbia_2019} investigated similar effects but considering dephasing coupling (no energy exchange between probe and bath). In \cite{Planella_2020} the authors analyse the thermometric consequences for of collective coupling between an ensemble of harmonic oscillators and the bath. Finally, in \cite{Ivanov_2019}, thermometry via collective spins is investigated. 
 However, the collective spin and the measured system (the collective centre-of-mass motion) are assumed to form a closed system so that the collective spin does not reach any steady state, and in particular the heat capacity does not play any role there. Still, it is shown in \cite{Ivanov_2019} that some collective effects can be beneficial outside of the low-temperature regime.   
 
For thermal machines, several works pointed at possible performance enhancements based on collective bath coupling. In \cite {Kloc_2019}, the suggested output power enhancement relies on equilibration speed-up stemming from collective effects in spins $1/2$.  
 Collective couplings have also been investigated in continuous thermal machines \cite{Niedenzu_2015,Niedenzu_2018,Watanabe_2019}. 
 Finally, in \cite{bathinducedcoh} it is shown that the combination of mitigation effects stemming from collective bath couplings can increase the output power of Otto engines. The present study confirms and extends the results of \cite{bathinducedcoh}.

In this paper, we exploit a common feature of thermal equilibrium thermometry and certain class of thermal machines: the central role played by the heat capacity of the probe or working medium. 
For thermometry, this was shown for instance in \cite{Correa_2015, Hofer_2019, Hovhannisyan_2018, Mehboudi_2019}. With respect to thermal machines, it was recently reported that Otto cycle thermal engines operating close to the Carnot efficiency \cite{Campisi_2016} have their output power determined by the heat capacity of the working medium. Remarkably, this was also proved to hold for some classes of non-ideal Carnot engines \cite{Abiuso_2019}, the so-called finite-time Carnot engines (working in the low-dissipative limit).
Taking advantage of this common characteristics, we study the impact of collective bath coupling on the heat capacity of spin ensembles and use these results to infer the corresponding effects on quantum thermometry and quantum engines. 

 We find that the heat capacity associated with an ensemble of spins interacting collectively with a thermal, called {\it collective heat capacity} in the following, can take value dramatically different from the sum of the individual heat capacities (which corresponds to situations where each spin interacts independently with the bath), called {\it independent heat capacity}. At hight temperature, 
 the collective heat capacity can become much larger than the independent heat capacity -- up to $(ns+1)/(s+1)$ times larger, where $n$ is the number of spins and $s$ their dimension. Conversely, at low temperature, the independent heat capacity is in general larger than the collective one -- up to $n$ times.  
  
 Applied to thermometry, we show that an ensemble of spins interacting collectively with the sample of interest can provide a higher precision for temperature estimation outside the very low-temperature regime. This is of interest for biological or medical applications, like intra-cellular thermometry \cite{Kucsko_2013, Baffou_2014}, in-vivo nanoscale thermometry \cite{Fujiwara_2020}, but also for nanoscale Chemistry \cite{Wu_2013} and thermal mapping of micro or nano scale electronic devices \cite{Mecklenburg_2015}. 
 We provide an approximated expression in terms of $n$ and $s$ of the critical temperature below which collective effects stop enhancing the precision of the temperature estimation. Applied to experimental data from \cite{Bouton_2020, Wang_2018, Liu_2019}, the critical temperature can indeed be very small, indicating that in practice, collective couplings can still enhance the temperature estimation precision over a very large range of sample's temperatures.
 
Finally, with respect to cyclic thermal machines -- Otto engines operating close to the Carnot bound, we show that collective effects are always beneficial in terms of output power, which is a stronger result than in \cite{bathinducedcoh}. The largest enhancements come at high hot bath temperatures. 
Regarding the output work per cycle, we recover the same asymptotic scaling as in \cite{bathinducedcoh}. 
 

 The paper is organised as follows. 
 In Section \ref{Experiment} we come back briefly on experimental realisations of spins collectively coupled to a thermal bath. In Section \ref{heatcapa} we derive some properties of the collective heat capacity for an ensemble of $n$ spins of arbitrary dimension $s$. In Sections \ref{thermo} and \ref{engines} we apply the results on collective heat capacity to quantum thermometry and cyclic quantum engines, respectively. We conclude in Section \ref{concl} with some final remarks and perspectives.



\section{About experimental realisations of collective coupling}\label{Experiment}
On the one hand, the philosophy behind this paper is to investigate a particular type of collective effect and analyse how beneficial it can be for certain thermodynamic tasks. Then, depending on the extent of the benefit, one can decide to start thinking of how to actually realise such collectively-enhanced devices. 
%
In this perspective, we briefly discuss in the following some possibilities for experimental realisations of collective coupling between spins and bath. 
The aim of this paper is to suggest that the benefits are worth the ``experimental battle''.

%

Ideally, we would think of adapting the aforementioned designs to include a spin ensemble collectively coupled to the bath. This is certainly possible in \cite{Lindenfels_2019} since therein the baths are emulated by an external magnetic field, offering the possibility of addressing collectively an ensemble of spins. While it might be possible to upgrade the other designs \cite{Peterson_2018, Wang_2018, Liu_2019, Bouton_2020} to collective bath coupling, it is less obvious than in \cite{Lindenfels_2019}. 

Beyond that, we stress that there are several known platforms realising collective coupling between a spin ensemble and electromagnetic modes \cite{Wood_2014, Angerer_2018, Barberena_2019, Tucker_2019} or even phonons \cite{Cernotik_2019, Ivanov_2019} and mechanical oscillators \cite{Karg_2020}. Note that spins commonly used in realisations of collective coupling are from atoms of strontium \cite{Barberena_2019,Tucker_2019,Norcia_2018} and rubidium \cite{Karg_2020}.  
Using such platforms, one can imagine the implementation of collective coupling between the spin ensemble and a bath or a sample of interest. Indeed, if the intermediary system -- the electromagnetic or phononic mode -- is coupled to the bath (or sample), the effective dynamics of the spin ensemble can be a collective dissipation. A required condition for that is having a coupling between the intermediary system and the bath larger than the coupling between the spin ensemble and the intermediary system. 
This is shown explicitly in \cite{Wood_2014} where the intermediary system is a cavity mode coupled to the external electromagnetic field playing the role of the thermal bath. This can be extended directly to other platforms since the core mechanism is the same, namely a spin ensemble interacting collectively with a bosonic mode which is itself interacting with a thermal bath.

The conclusion of this section is that collective coupling of a spin ensemble with a thermal bath is tangible in several platforms, and even readily realisable in the experimental design used in \cite{Angerer_2018, Barberena_2019, Tucker_2019}. 

\section{Collective heat capacities}\label{heatcapa}
 The heat capacity $C$ of a system in a thermal state at temperature $T$ determines how much energy must be absorbed (or released) to increase (or decrease) the system's temperature by an amount $\delta T$. It is naturally given by $C= \frac{\partial E^{\rm th}(\beta)}{\partial T}=-k_B\beta^2 \frac{\partial E^{\rm th}(\beta)}{\partial \beta}$, where $E^{\rm th}(\beta)={\rm Tr} H \rho^{\rm th}(\beta)$ is the energy of the system in the thermal state $ \rho^{\rm th}(\beta):=Z^{-1}(\beta) e^{-\beta H}$ at inverse temperature $\beta:=1/k_BT$, $k_B$ is the Boltzmann constant, $Z(\beta)={\rm Tr} e^{-\beta H}$ is the partition function and $H$ is the free Hamiltonian of the system. As one could expect, the heat capacity plays a central role in thermometry and thermal machines. We will come back on this aspect in Sections \ref{thermo} and \ref{engines}. For now we focus on the heat capacity of an ensemble of spins interacting collectively with a thermal bath. The main idea is that since the steady state energy of a spin ensemble interacting collectively with a thermal bath \cite{bathinducedcoh,bathinducedcohTLS} is different from the thermal equilibrium energy -- reached when all spins interact independently with the bath, the {\it collective} and {\it independent} heat capacities should also be different. 
 
 More precisely, we consider an ensemble of $n$ identical spins of dimension $s$ and free Hamiltonian $H = \hbar \omega J_z$, where $J_z := \sum_{k=1}^n j_{z,k}$ is the sum of the z-component of the local angular momentum operators associated to each spin $k$ (similar notations for the $x$ and $y$ components).  The collective coupling with a thermal bath corresponds to a coupling Hamiltonian of the form $V=g J_x O_B$, where $O_B$ is an unspecified bath observable, and $g$ represents the coupling strength. 
 Under the usual Born, Markov, and secular approximations \cite{Petruccione_Book, Cohen_Book}, the dynamics of the collective dissipation is of the form \cite{bathinducedcoh}
\bea\label{colme}
\frac{ d \rho}{dt} &=& \Gamma(\omega) \left(J^{-}\rho J^{+} - J^{+} J^- \rho\right) \nn\\
&&+ \Gamma(-\omega) \left(J^{+}\rho J^{-} - J^{-} J^{+} \rho\right)  + {\rm h.c.},
\eea
where $J^{\pm}:= J_x \pm iJ_y$ are the collective jump operators of the spin ensemble, $\Gamma(\omega) = \hbar^2g^2\int_0^{\infty} e^{i\omega u} {\rm Tr}\rho_B O_B(u)O_B du$ is the ``half Fourier transform'' of the bath correlation function, $\rho_B$ is the density operator of the thermal bath at temperature $T$, and $O_B(u)$ denotes the interaction picture of $O_B$. 
 The steady state of the above collective dissipation \eqref{colme} can be expressed in a relative simple way using the {\it collective basis} $|J,m\rangle_i$ \cite{Sakurai_Book} made of the common eigenvectors of $J_z$ and ${\cal J}^2:=J_x^2+J_y^2+J_z^2$,
 \bea
&&{\cal J}^2|J,m\ket_i = \hbar J(J+1)|J,m\ket_i\nn\\
&& J_z|J,m\ket_i= \hbar m|J,m\ket_i,
\eea
 with $-J\leq m\leq J$ and $J \in [J_0; ns]$, where $J_0 =0$ if $s\geq 1$ and $J_0=1/2$ if $s=1/2$ and $n$ odd. The index $i$ belongs to the interval $[1;l_J]$, where $l_J$ denotes the multiplicity of the eigenspaces associated to the eigenvalue $J$ of the operator ${\cal J}^2$. With these notations, the steady state takes the form \cite{bathinducedcoh}
\be\label{colss}
\rho^{\rm ss} (\beta) = \sum_{J=J_0}^{ns} \sum_{i=1}^{l_J}p_{J,i}\rho_{J,i}^{\rm th}(\beta)
\ee
where $p_{J,i}:=\sum_{m=-J}^J~_i\bra J,m|\rho_0|J,m\ket_i$ is the weight of the initial state of the spin ensemble $\rho_0$ in the eigenspace of total spin $J$ and \be\label{thstji}
\rho_{J,i}^{\rm th}(\beta):=Z_{J}(\beta)^{-1}\sum_{m=-J}^J e^{- m\hbar\omega\beta}|J,m\ket_i\bra J,m|,
\ee
 with $Z_J(\beta):=\sum_{m=-J}^J e^{-m\hbar\omega\beta}$.
 Note that if the initial state contains some coherences of the type $_i\langle J,m|\rho_0|J,m\rangle_{i'}$, with $i\ne i'$, it is not proven that the corresponding steady state has exactly the form \eqref{colss} (see \cite{bathinducedcoh}). A short note is provided in Appendix \ref{note} regarding the stability of the collective steady state under the emergence of small spin-spin interactions or small disorder and inhomogeneities altering the local energy levels of each spin.
 Additionally, the fact that the steady state retains some dependence on the initial state is a consequence of a dynamics with multiple steady states. The steady state actually reached is determined by the initial conditions \cite{Merkli_2015, bathinducedcoh, bathinducedcohTLS}. This should not be seen as a contradiction with the Markovian dynamics. In particular, one can verify that in the present dynamics there is not backflow of information as it typically occurs in non-Markovian dynamics.  
 

The energy of the spin ensemble when it reaches the steady state \eqref{colss} is 
\bea\label{colen}
E^{\rm ss}(\beta) &:=& \hbar\omega {\rm Tr} J_z \rho^{\rm ss}(\beta) \nn\\
&=& \sum_{J=J_0}^{ns}\sum_{i=1}^{l_J} p_{J,i}e_J(\beta),
\eea 
with $e_J(\beta):=\hbar\omega {\rm Tr} J_z \rho_{J,i}^{\rm th}(\beta)= \hbar\omega \sum_{m=-J}^J m \frac{e^{-m\hbar\omega \beta}}{Z_J(\beta)}$.
%
Then, quite naturally, we can define the {\it collective} heat capacity as the derivative with respect to the bath temperature of the steady state energy reached via collective dissipation, namely
\bea\label{cindist}
C^{\rm col}(\beta)&:=& -k_B\beta^2 \frac{\partial E^{\rm ss}(\beta)}{\partial \beta} \nn\\
&=& \sum_{J=J_0}^{ns} \sum_{i=1}^{l_J}p_{J,i}C_J(\beta),
\eea
with 
\bea\label{CJ}
C_J(\beta)&:=& -k_B\beta^2 \frac{\partial e_J(\beta)}{\partial \beta} \nn\\
&=& k_Bb^2\left[ \left(\frac{1/2}{\sinh b/2}\right)^2-\left(\frac{J+1/2}{\sinh(J+1/2)b}\right)^2\right],\nn\\
\eea
where $b:=\hbar\omega \beta$. \\

One can verify that $C_J(\beta)>C_{J'}(\beta)$ for $J>J'$ and for all $\beta$, even for negative effective bath temperature -- relevant in some specific situations like in presence of several thermal baths \cite{Brunner_2012,autonomousmachines} or spin baths \cite{Assis_2018,Kosloff_2019}. This implies that the largest collective heat capacity is obtained with initial state such that $p_{J=ns} =1$. Such states span a subspace sometimes called the  Dick or symmetrical subspace. In particular, some experimentally simple states like thermal states at inverse temperature $|\beta_0| \gg 1/\hbar\omega$ belong to such subspace.
 As expected, the applications to thermometry and thermal engines seek the largest heat capacity. Therefore, now we know that the largest advantage obtained from collective interactions is achieved for initial state belonging to the symmetrical subspace. In the following we compare the best case scenario, $C_+^{\rm col}(\beta):= C_{J=ns}(\beta)$, to the {\it independent} heat capacity.  \\

\begin{figure}
\centering
(a)\includegraphics[width=6cm, height=4cm]{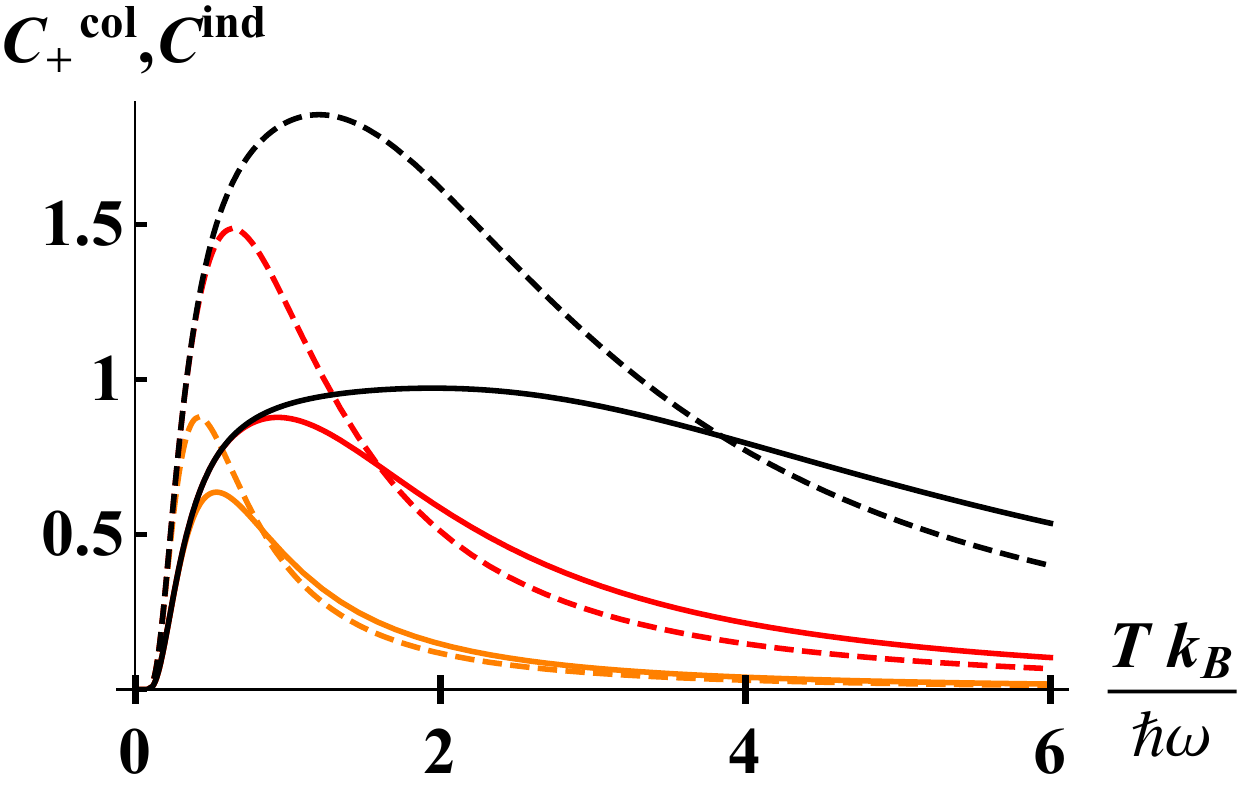}\\
(b)\includegraphics[width=6cm, height=4cm]{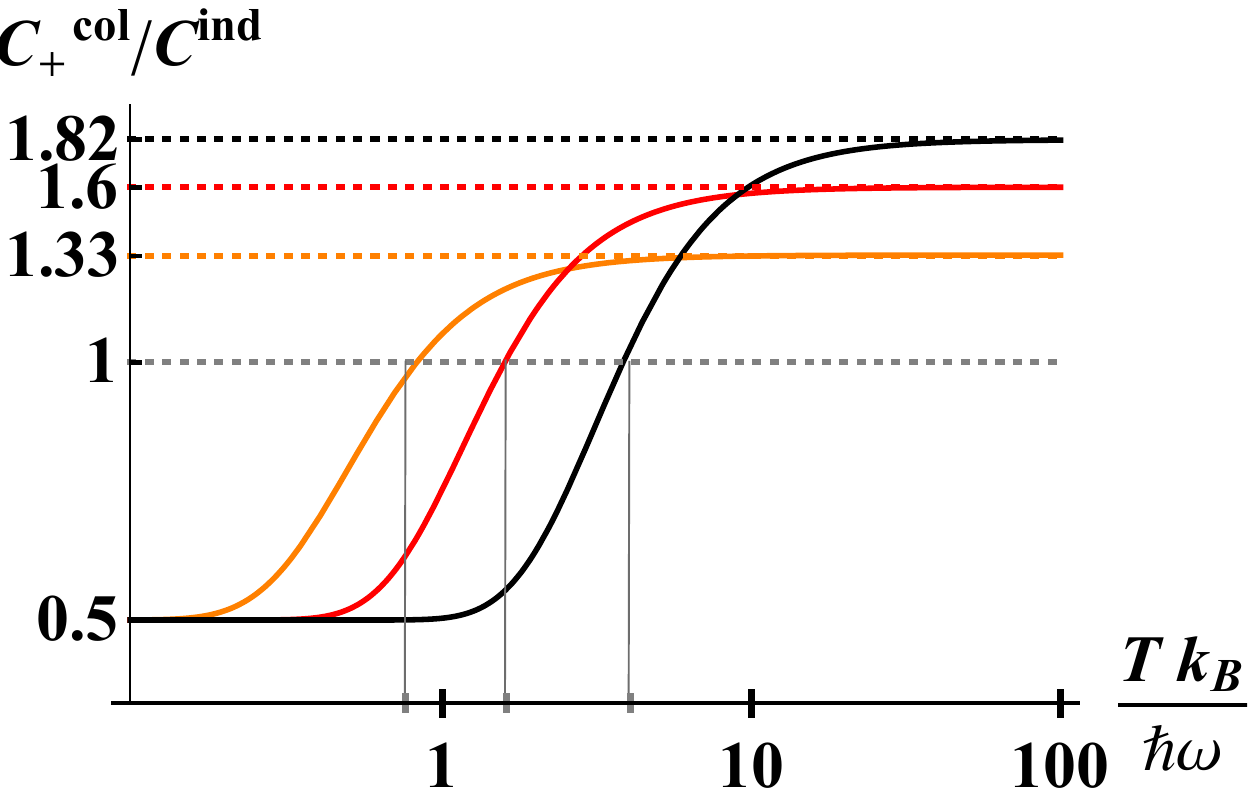}
\caption{(a) Heat capacities of the collective spins $C_{+}^{\rm col}(\beta)$ (solid lines) and of the independent spins $C^{\rm ind}(\beta)$ (dashed lines) as functions of $k_BT/\hbar\omega$ for ensembles of $n=2$ spins of dimension $s=1/2$ (orange curves), $s=3/2$ (red curves), $s=9/2$ (black curves). (b) Ratios of the heat capacities $C_{+}^{\rm col}(\beta)/C^{\rm ind}(\beta)$ (in log-log) for the same values of $n$ and $s$. The dotted lines indicate the asymptotic behaviours which follow the analytical value $(ns+1)/(s+1)$. The vertical gray lines indicates the critical temperature $T_{\rm cr}(n,s)$ given by the approximate expression \eqref{tcr}. }
\label{capacityspin}
\end{figure}

\begin{figure}
\centering
(a)\includegraphics[width=6cm, height=4cm]{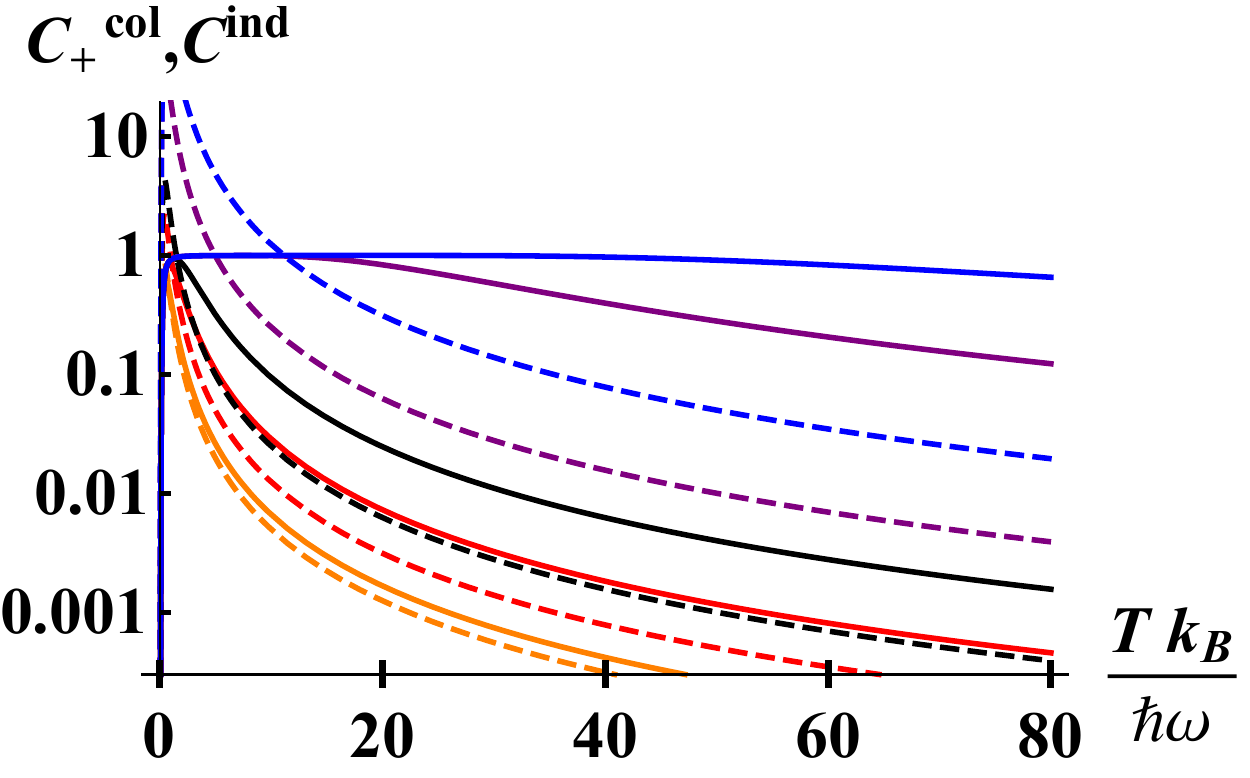}
(b)\includegraphics[width=8cm, height=5.5cm]{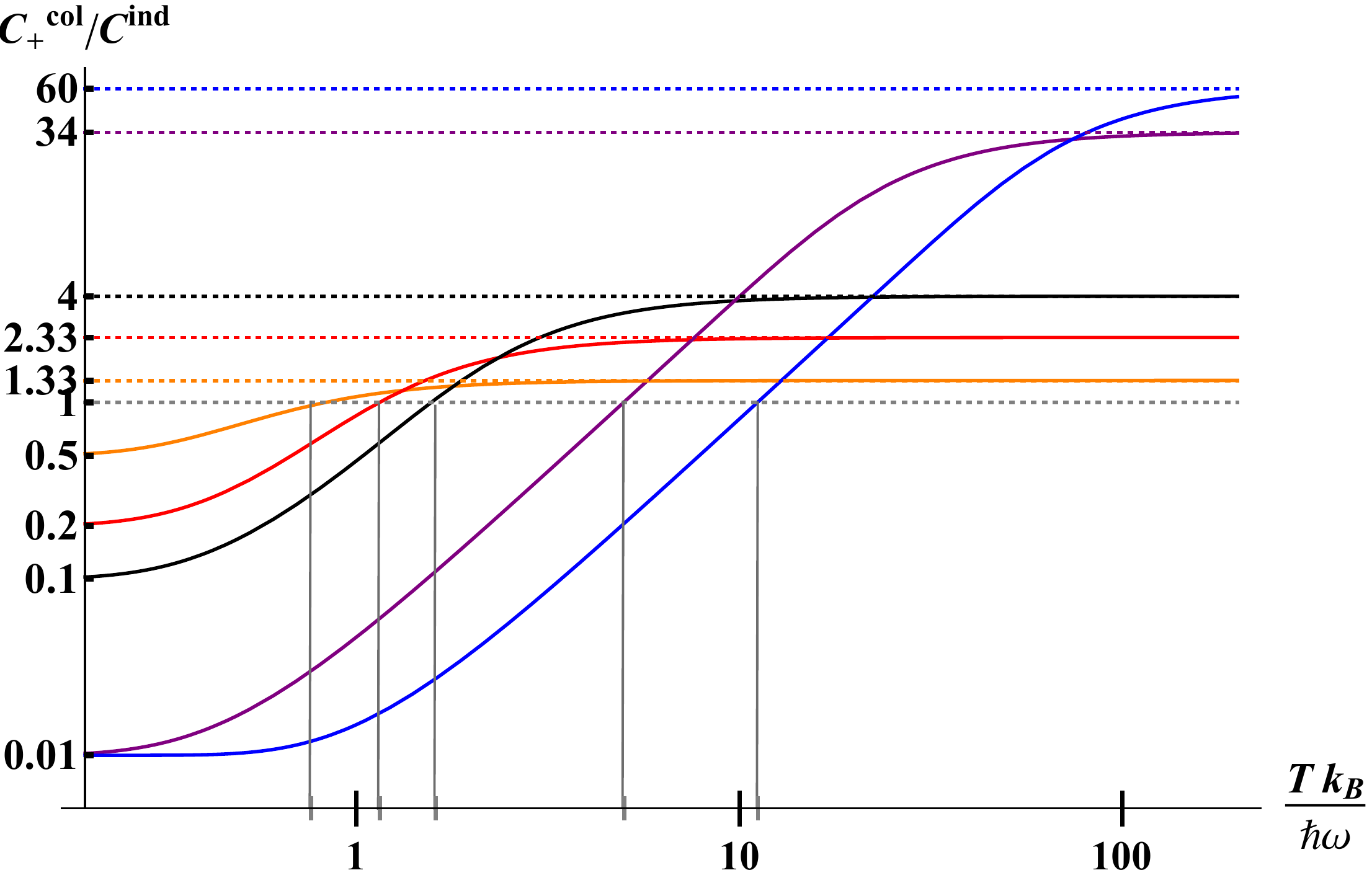}
\caption{(a) Heat capacities of the collective spins $C_{+}^{\rm col}(\beta)$ (solid lines) and of the independent spins $C^{\rm ind}(\beta)$ (dashed lines) as functions of $k_BT/\hbar\omega$ (semi-log scale) for ensembles containing $n=2$ (orange curve), $n=5$ (red curve), $n=10$ (black curve), $n=100$ (purple curve) spins $s=1/2$. The blue curve corresponds to an ensemble of $n=100$ spins $s=3/2$. (b) Ratios of the heat capacities $C_{+}^{\rm col}(\beta)/C^{\rm ind}(\beta)$ (in log-log) as a function of $k_BT/\hbar\omega$. The colour code is the same as in the panel (a). The dotted lines indicate the asymptotic behaviours which follow the analytical value $(ns+1)/(s+1)$. The vertical gray lines indicates the critical temperature $T_{\rm cr}(n,s)$ given by the approximate expression \eqref{tcr}.}
\label{capacityn}
\end{figure}

{\it Comparison with independent heat capacity}. The {\it independent} heat capacity is the derivative with respect to the bath temperature of the thermal equilibrium energy $E^{\rm th}(\beta)$ -- the energy reached when each spin interacts independently with the bath,
\bea\label{cth}
C^{\rm ind}(\beta)&:=& -k_B\beta^2 \frac{\partial E^{\rm th}(\beta)}{\partial \beta} \nn\\
&=& n C_{J=s}(\beta),
\eea
where 
\be\label{energyth}
E^{\rm th}(\beta) := \hbar\omega{\rm Tr} J_z \rho^{\rm th}(\beta) = n e_{J=s}(\beta),
\ee
 $\rho^{\rm th}(\beta) = Z(\beta)^{-1} e^{-\hbar\omega\beta J_z}$, and $Z(\beta) = {\rm Tr} e^{-\hbar\omega\beta J_z}$.\\
Then, we are left to compare $C_+^{\rm col}(\beta)= C_{J=ns}(\beta)$ and $C^{\rm ind}(\beta) = nC_{J=s}(\beta)$. The expansion of the expression \eqref{CJ} at $\hbar \omega |\beta| \gg 1$ gives respectively
$C_+^{\rm col}(\beta) \underset{\hbar\omega |\beta|\gg 1}{\sim} (\hbar\omega \beta)^2e^{-\hbar\omega |\beta|}$,
and $C^{\rm ind}(\beta) \underset{\hbar\omega |\beta|\gg 1}{\sim}  n(\hbar\omega \beta)^2e^{-\hbar\omega |\beta|}$.
In particular, 
\be
C_+^{\rm col}(\beta) /C^{\rm ind}(\beta)  \underset{\hbar\omega |\beta|\gg 1}{\sim} n^{-1}.
\ee
 By contrast, for $\hbar\omega |\beta|\ll 1$, we obtain
\be\label{ascol}
C_+^{\rm col}(\beta) = \frac{1}{3} ns(ns+1)(\hbar\omega \beta)^2 + {\cal O}\big[(ns\hbar\omega \beta)^4\big],
\ee
and
\be
C^{\rm ind}(\beta) = \frac{n}{3} s(s+1)(\hbar\omega \beta)^2 + {\cal O}\big[n(s\hbar\omega \beta)^4\big],
\ee
implying $C_+^{\rm col}(\beta) /C^{\rm ind}(\beta) \underset{\hbar\omega |\beta|\ll 1}{=} \frac{ns+1}{s+1} +{\cal O}\big[n(n\hbar\omega \beta)^2\big]$.
For the intermediary regime between these asymptotic limits, the behaviour of $C_+^{\rm col}(\beta)$,  $C^{\rm ind}(\beta)$, and  $C_+^{\rm col}(\beta)/C^{\rm ind}(\beta)$ is represented in Figs. \ref{capacityspin} and \ref{capacityn}
as functions of $k_BT/\hbar\omega$ for several value of $n$ and $s$.

Importantly, the critical value $T_{\rm cr}$ of the bath temperature such that $C_+^{\rm col}(1/k_BT_{\rm cr}) = C^{\rm ind}(1/k_BT_{\rm cr})$ is well approximated by the function 
\be\label{tcr}
\frac{k_BT_{\rm cr}(n,s)}{\hbar\omega} \simeq \left(\frac{4ns(s+1)+1}{12}\right)^{1/2}.
\ee
A comparison with numerical solutions gives very good agreement, see Figs. \ref{capacityspin} (b) and \ref{capacityn} (b). The above expression \eqref{tcr} was obtained expanding the expressions of $C^{\rm ind}(\beta)$ and $C_{+}^{\rm col}(\beta)$ using the assumptions that $\hbar\omega/k_BT_{\rm cr} \ll 1$ and $(ns+1/2)\hbar\omega/k_BT_{\rm cr} \gg 1$ for growing $n$. Such assumptions are based on observations of the numerical solutions.

\section{Applications to quantum thermometry}\label{thermo}
We consider the situation where the sample we want to estimate the temperature is much larger than the probe -- our spin ensemble. Under weak coupling, one can consider that the spin ensemble reach a steady state without indeed affecting the sample. In other words, the sample plays the role of a thermal bath. This is the general framework considered by quantum thermometry in thermal equilibrium \cite{Mehboudi_2019}. Additionally to this framework, and accordingly to the above study of heat capacity, we consider two different situations. In what we call the independent-dissipation situation, each spin of the ensemble interacts independently with the sample so that the spin ensemble eventually reaches a thermal state at the sample's temperature $T$. In the other situation, the collective-dissipation situation, the spin interact collectively with the sample and reaches the steady state $\rho^{\rm ss}(\beta)$ of Eq. \eqref{colss}. 

The maximal information extractable from the spin ensemble about the sample's temperature is usually quantified by the quantum Fisher information ${\cal F}(T)$ \cite{Braustein_1994, Helstrom_1976, Holevo_1982}. Quite intuitively, the maximal precision of the estimation of the sample's temperature can be related to this maximal extractable information ${\cal F}(T)$. This is indeed established by the Cramer-Rao bound \cite{CramerRao}. Therefore, ${\cal F}(T)$ directly informs about the maximal achievable precision of the estimation of the sample's temperature. As mentioned in the introduction, it was shown in \cite{Correa_2015, Hofer_2019, Hovhannisyan_2018, Mehboudi_2019} that 
${\cal F}(T) = C(T)/k_BT^2 = \Delta^2 \langle H\rangle/T^2$, where $C(T)$ is the heat capacity of the probe and $\Delta^2  \langle H\rangle$ is its energy variance. 
However, such property is valid when the probe is in a thermal state, and in principle not valid for non-thermal states. Therefore, before applying the results of the last section regarding collective heat capacity we have to show that the maximum precision for the temperature estimation using non-thermal states of the form $\rho^{\rm ss}(\beta)$ is indeed given by the collective heat capacity. This is shown in Appendix \ref{appcolqfi} where we establish in particular that the quantum Fisher information for any state of the form $\rho^{\rm ss}(\beta)$ is 
\be\label{qfimain}
{\cal F}^{\rm col}(T) = C^{\rm col}(T)/k_BT^2.
\ee
 Then, it follows from the analysis of the last section that the largest enhancements from collective coupling are obtained for spin ensembles initially in a thermal state at very large inverse temperature $|\beta_0|\ll1/\hbar\omega$, or more generally for initial states belonging to the symmetrical subspace. We denote by ${\cal F}^{\rm col}_{+}(T)$ the corresponding quantum Fisher information. This is to be compared with the quantum Fisher information ${\cal F}^{\rm ind}(T)$ obtained from independent dissipation of each spin, which is equal to ${\cal F}^{\rm ind}(T) = C^{\rm ind}(T)/k_BT^2$ since the steady state in this case is a thermal state. 
 Then, the ratio ${\cal F}^{\rm col}_{+}(T)/{\cal F}^{\rm ind}(T)$ is equal to $C^{\rm col}_{+}(\beta)/C^{\rm ind}(\beta)$ which is represented for some values of $n$ and $s$ in Fig. \ref{capacityspin} (b) and Fig. \ref{capacityn} (b). 
 
\begin{figure}
\centering
(a)\includegraphics[width=6cm, height=4cm]{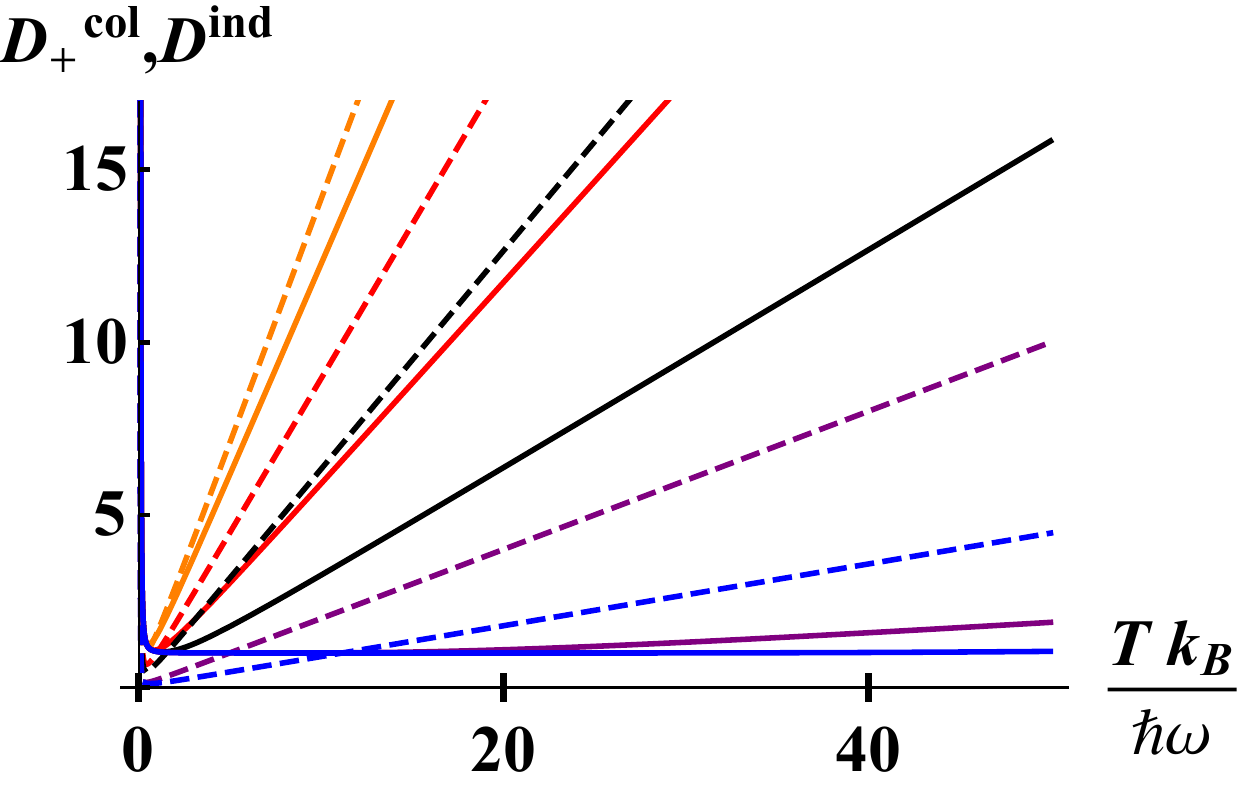}\\
(b)\includegraphics[width=8cm, height=5.5cm]{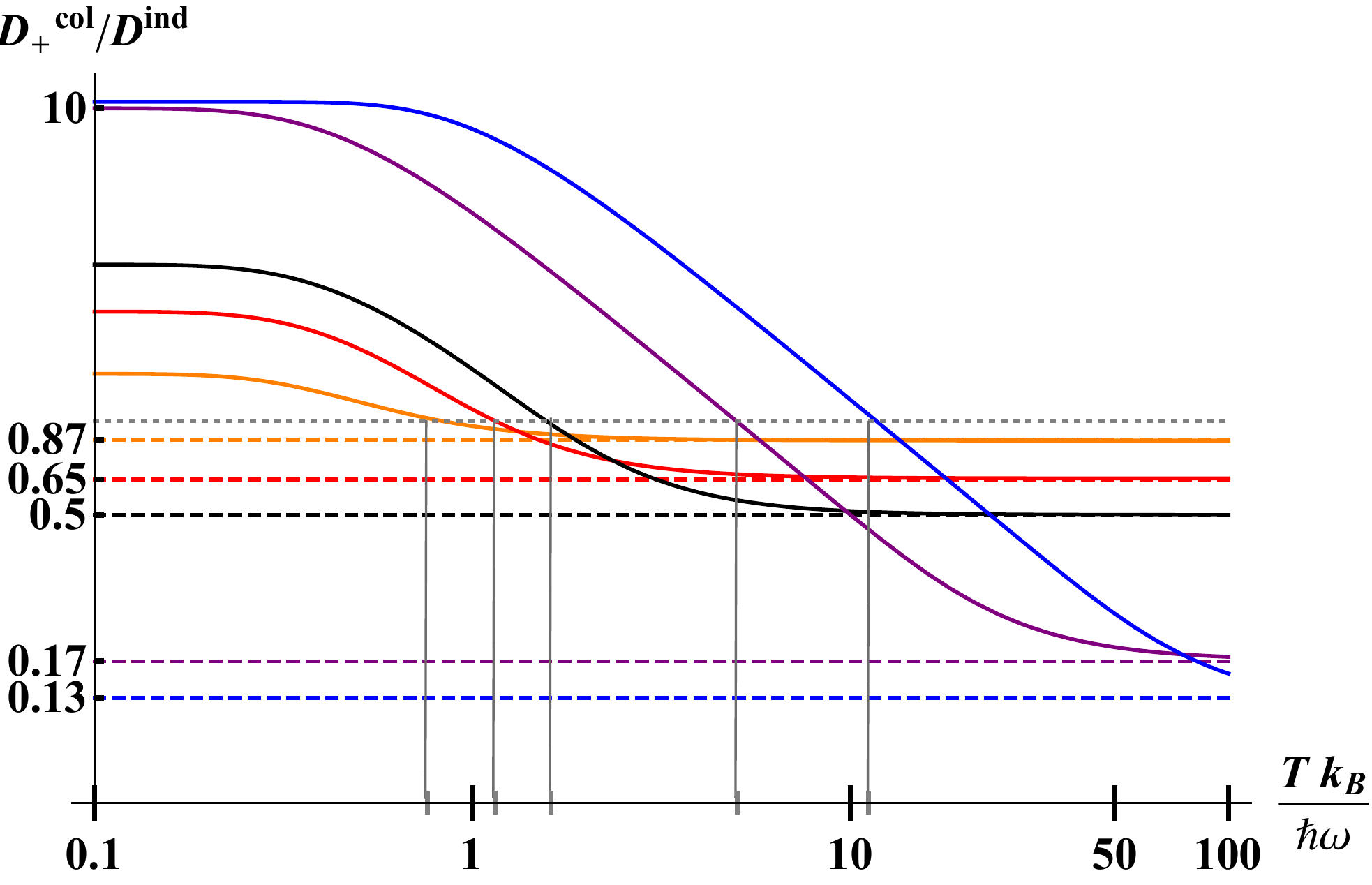}
\caption{(a) Minimal relative standard deviations $\left(\frac{\Delta T}{T}\right)_{\rm min}$, denoted by $D_{+}^{\rm col}$ and $D^{\rm ind}$, in function of the sample's temperature $\hbar\omega T$
and for ensembles containing $n=2$ (orange curve), $n=5$ (red curve), $n=10$ (black curve), $n=100$ (purple curve) spins $s=1/2$. The blue curve represents an ensemble of $n=100$ spins $s=3/2$. The solid curves correspond to collective interactions with the sample when the spin ensemble is initialised in the symmetrical subspace. The dashed curves represent the precision obtained through independent coupling with the sample. (b) Ratios of the minimal relative standard deviations $D_{+}^{\rm col}/D^{\rm ind}$. The color code is the same as in the panel (a). The vertical gray lines indicate the critical temperatures $T_{\rm cr}(n,s)$ as estimated by Eq. \eqref{tcr}. The horizontal dotted gray line indicates 1 as guide for the eyes. The dashed lines indicate the asymptotic behaviours which follow the analytical value $\sqrt{(s + 1)/(ns + 1)}$.
 }
\label{minimalv}
\end{figure}

 In terms of the relative precision $\Delta T/T$, where $\Delta T$ represents the standard deviation\footnote{More precisely, each estimation, obtained after $\nu$ measurements, is a random variable whose distribution has a standard deviation denoted by $\Delta T$. The smaller is $\Delta T$, the more precise is the estimation process. } of the estimated value of $T$, we have
\be\label{maxprecision}
\frac{\Delta T}{T} \geq \nu^{-1/2}\left(\frac{\Delta T}{T}\right)_{\rm min} := \frac{1}{\sqrt{\nu C^{\rm col}_{+}(T)/k_B}}
\ee
 where $\nu$ is the number of measurements used to realise one estimation of $T$. The above inequality can be saturated for instance when choosing the maximum likelihood estimator. Importantly, for large sample's temperature $k_BT\gg\hbar\omega$ one obtains from \eqref{ascol} an Heisenberg scaling \cite{Mehboudi_2019}: $\Delta T \sim 1/n$.
 Fig. \ref{minimalv} (a) presents the plots of the minimal relative standard deviation $\left(\frac{\Delta T}{T}\right)_{\rm min}$ for collective and independent coupling with the sample, denoted respectively by $D_{+}^{\rm col}$ and $D^{\rm ind}$, in function of the sample's temperature. Fig. \ref{minimalv} (b) provides the ratio $D_{+}^{\rm col}/D^{\rm ind}$ of the minimal relative variances. We have the following asymptotic scaling for large sample's temperature $D_{+}^{\rm col}/D^{\rm ind} \underset{k_BT\gg \hbar\omega}{\sim}  \sqrt{(s+1)/(ns+1)}$.

 In particular, the temperature of the sample can be estimated with a higher precision thanks to collective interaction as long as $T > T_{\rm cr}(n,s)$.
 To have an estimate of what would be the critical temperature we can use experimental data from \cite{Bouton_2020}. The energy splitting of the cesium quasispin is about $\hbar\omega \sim 2.4 \times 10^{-30} J$. It means that for an ensemble of $n=2$ cesium atoms, collective interaction become advantageous for sample's temperature of the order or superior to $T_{\rm cr}(n=2,s=7/2) \simeq 5.5 {\rm nK}$ (the cesium quasispin is of dimension $s=7/2$), and increasing the precision by a factor up to $1.8$. For larger ensembles of for instance $n=10$, the critical temperature is $T_{\rm cr}(n=10,s=7/2) \simeq 12 {\rm nK}$ and collective interactions can increasing the precision by a factor 8. 
 For the NV-centre spins used in \cite{Wang_2018, Liu_2019}, the energy splitting is of the order of $\hbar\omega \simeq 1.9\times 10^{-24} J$, implying that collective interactions become advantageous from $T_{\rm cr}(n=10, s=1/2) \simeq 0.22 {\rm K}$ for an ensemble of $n=10$ NV center spins, with precision increased by a factor up to 4.


 Finally, there are two interesting remarks to be made around the proof of Eq.\eqref{qfimain}. First, while the maximum precision is given by the collective heat capacity $C^{\rm col}(T)$, it is not equal to the variance of the energy unlike thermal states. In fact, the collective heat capacity is in general strictly smaller than the energy variance $\Delta^2 \langle H\rangle$. Secondly, the best measurement is not the energy measurement but a non-local measurements corresponding to projections onto the collective states $\{|J,m\rangle_i\}_{J_0\leq J\leq ns, -J\leq m\leq J, 1\leq i\leq l_J}$. Since such measurements sounds experimentally unrealistic, we should consider only the precision yielded by local energy measurements. Fortunately, in the particular situation where the initial state of the spin ensemble belongs to the symmetrical subspace, local energy measurements turn out to be optimal. The conclusion of Appendix \ref{appcolqfi} is that when considering the best case scenario -- the spin ensemble initially in the symmetrical subspace -- the maximal precision of the temperature estimation
 can be reached by usual local energy measurements and is given by the collective heat capacity.

%

\section{Applications to quantum engines}\label{engines}
\subsubsection{Work per cycle near Carnot efficiency}
In this section we consider a quantum engine operating according to the quantum Otto cycle \cite{Scully_2002,Quan_2007}. The working medium is an ensemble of $n$ spins of dimension $s$ of Hamiltonian $H(\lambda_t) = \lambda_t \hbar \omega J_z$, where $\lambda_t$ is the compression factor which varies continuously between $\lambda_c$ and $\lambda_h$ during the isentropic strokes. The two isochoric strokes realised alternatively in contact with hot and cold baths bring the spin ensemble to the usual thermal equilibrium state $\rho^{\rm th}(T_x,\lambda_x)=Z^{-1}(T_x,\lambda_x)e^{-H(\lambda_x)/k_B T_x}$, with $x=c,h$ and $Z(T_x,\lambda_x) = {\rm Tr}e^{-H(\lambda_x)/k_B T_x}$ if each spin interacts independently with the bath. However, if the spins interact collectively with the successive bath, the isochoric strokes result in the steady state $\rho^{\rm ss}(T_x,\lambda_x)= \sum_{J=J_0}^{ns} \sum_{i=1}^{l_J}p_{J,i}\rho_{J,i}^{\rm th}(T_x,\lambda_x)$,
where 
 \be
\rho_{J,i}^{\rm th}(T_x,\lambda_x):=\sum_{m=-J}^J \frac{e^{- m\lambda_x\hbar\omega/k_BT_x}}{Z_{J}(T_x,\lambda_x)}|J,m\ket_i\bra J,m|,
\ee
 with $Z_J(T_x,\lambda_x):=\sum_{m=-J}^J e^{-m\lambda_x\hbar\omega/k_BT_x}$.
Importantly, the weight $p_{J,i}=\sum_{m=-J}^J~_i\bra J,m|\rho_0|J,m\ket_i$ in each eigenspace of total spin $J$ is constant throughout the cycles \cite{bathinducedcoh} and is determined by $\rho_0$, the state of the spin ensemble before the engine is switched on.


 Then, for a Otto cycle operating close to the Carnot efficiency, the extracted work per cycle is \cite{Campisi_2016} (see also a brief derivation in Appendix \ref{wpercycle})  
\bea
W &=&\Delta \eta \lambda_h^2 (\beta_c-\beta_h)\frac{C(\theta_h)}{k_B\theta_h^2} + {\cal O}(\Delta \eta^2)
\eea
where, $\Delta \eta = \eta_c-\eta=\frac{\lambda_c}{\lambda_h}-\frac{\beta_h}{\beta_c}$ is the difference between the Carnot efficiency $\eta_c:=1-\frac{\beta_h}{\beta_c}$ and the actual efficiency, $\theta_x:=\lambda_x\beta_x$ and $C(\theta_h)$ denotes generically the collective or independent heat capacity depending whether the spins interact collectively or independently with the baths.

\begin{figure}
\centering
(a)\includegraphics[width=7cm, height=4.5cm]{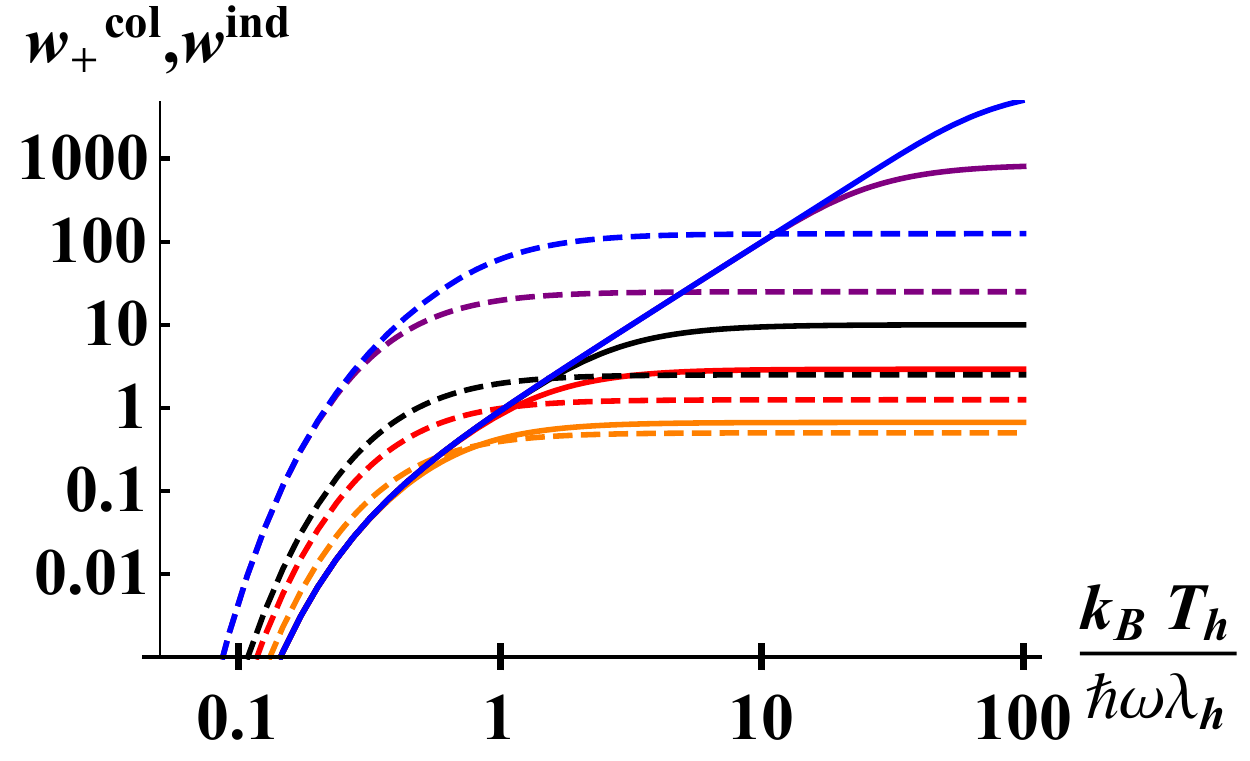}\\
\caption{Plots of $w^{\rm col}_{+}:= W^{\rm col}_{+}/(\Delta \eta \lambda_h^2\Delta\beta)$ (solid curves) and $w^{\rm ind}:=W^{\rm ind}/(\Delta \eta \lambda_h^2\Delta\beta)$ (dashed curves) as a function of $k_BT_h/\hbar\omega \lambda_h$ for fixed $\Delta \eta$, $\lambda_h$, $\Delta \beta$, and for ensembles containing $n=2$ (orange curve), $n=5$ (red curve), $n=10$ (black curve), $n=100$ (purple curve) spins $s=1/2$. The blue curves represent an ensemble of $n=100$ spins $s=3/2$.} 
\label{work}
\end{figure}

 The central question is what are the parameters yielding the largest output work and which of the independent or collective spin machine gives the largest work per cycle? Considering the best case scenario for the collective spin machine, meaning that $\rho_0$ belongs to the symmetrical subspace, we have to compare $W^{\rm col}_+ :=\Delta \eta \lambda_h^2 (\beta_c-\beta_h)\frac{C_{J=ns}(\theta_h)}{k_B\theta_h^2} + {\cal O}(\Delta \eta^2)$ with $W^{\rm ind} :=\Delta \eta \lambda_h^2 (\beta_c-\beta_h)n\frac{C_{J=s}(\theta_h)}{k_B\theta_h^2} + {\cal O}(\Delta \eta^2)$.
We are looking for the parameters maximising the output work at constant efficiency. Then, as expected, we find that the larger $\Delta \beta:=\beta_c-\beta_h$, the larger the output work. Considering now $\Delta\beta$ fixed, we are left with two parameters, $\lambda_h$ and $\theta_h$. Fixing firstly $\lambda_h$ (one can verify that $\theta_h$ can be changed while keeping $\Delta \eta$, $\Delta \beta$ and $\lambda_h$ fixed), the best choice is taking $\beta_h$ to zero. This is because $C_J(\theta)/\theta^2$ is monotonic decreasing (even though $C_J(\theta)$ is not monotonic, see for instance Fig. \ref{capacityspin}).
Since the maximum of $\frac{C_{J}(\theta)}{\theta^2}$ is $\frac{\hbar^2\omega^2}{12}[(2J+1)^2-1]$, we obtain for all $\lambda_h$, $\Delta \eta$ (to the first order), and $\Delta\beta$, 
 \be
 W^{\rm ind} \leq W^{\rm ind}_{\rm max}:=\Delta \eta \lambda_h^2\beta_c \frac{\hbar^2\omega^2}{12}n[(2s+1)^2-1],
 \ee
 and 
 \be
 W^{\rm col}_{+} \leq W^{\rm col}_{\rm max}:=\Delta \eta \lambda_h^2\beta_c \frac{\hbar^2\omega^2}{12}[(2ns+1)^2-1].
 \ee
 Note that both maximal values are reached for $\hbar\omega\beta_h$ going to 0 and that $W^{\rm col}_{\rm max} = \frac{sn+1}{s+1}W^{\rm ind}_{\rm max}$. We recover the asymptotic result of \cite{bathinducedcoh}. For intermediary value of $\hbar\omega\beta_h$, the plots of $w^{\rm col}_{+}:= W^{\rm col}_{+}/(\Delta \eta \lambda_h^2\Delta\beta)=C^{\rm col}_{+}(\theta_h)/k_B\theta_h^2 $ and $w^{\rm ind}:=W^{\rm ind}/(\Delta \eta \lambda_h^2\Delta\beta)=C^{\rm ind}(\theta_h)/k_B\theta_h^2$ are given in Fig. \ref{work} for ensembles of $n=2$ to $n=100$ spins.

  {\it Observation about reaching asymptotically the Carnot bound at finite power--.}  
In \cite{Campisi_2016, Abiuso_2019} the authors show that one could in principle reach asymptotically the Carnot bound while having a non-zero output power. One possibility is for instance to use phase transitions where the heat capacity scales up super-linearly in the number of atoms/subsystems in the working medium. Then, naturally, one could think that the collective effects on heat capacity shown here could be also useful for such purposes. This is in fact not the case as we explain in the following. 

For finite value of $\hbar\omega\beta_h$ and for arbitrary $n$, the amplification is upper bounded by $W^{\rm col}\leq (\hbar\omega)^2 \Delta \eta \lambda_h^2 (\beta_c-\beta_h) \left(\frac{1/2}{{\rm sinh}\hbar\omega\theta_h/2}\right)^2$. In other words, if we consider $\beta_h \ne 0$ (as in realistic conditions) and fixed, increasing the size of the working medium will increase the output power only until it reaches the above saturation limit. Therefore, it cannot be helpful to reach asymptotically the Carnot bound at finite power (where the idea is to take advantage of a super-linear scaling in the power per cycle to increase slowly the efficiency). 
  
  The above observation emphasises a drawback. At finite $T_h$, there is always a critical number $n_{\rm cr}$ of spins such that for spin ensembles larger than $n_{\rm cr}$ the independent-spin engine performs better than the collective-spin one. One can estimate the critical spin number from \eqref{tcr} and obtains $n_{\rm cr}(T_h,\lambda_h,s) \simeq \frac{3(T_h/\hbar\omega\lambda_h)^2-1/4}{s(s+1)}$.
  In the same spirit, for fixed bath temperatures, there is always a critical value $\lambda_{h, \rm cr}$ of the compression factor above which the collective-spin engine become less performant than the independent-spin engine. From \eqref{tcr} we have $\lambda_{h, \rm cr}(T_h,n,s)\simeq T_h\sqrt{12}/\left[\hbar\omega\sqrt{4ns(s+1)+1}\right]$. 
   On the other hand, the range of compression factors is always experimentally limited so that for finite $n$ -- since $n$ is also experimentally limited -- the collective spin engine performs always better that the independent one as soon as $T_h/\lambda_h\geq T_{\rm cr}(n,s)$.


\subsubsection{What about the output power?}
\begin{figure}
\centering
(a)\includegraphics[width=7cm, height=4.5cm]{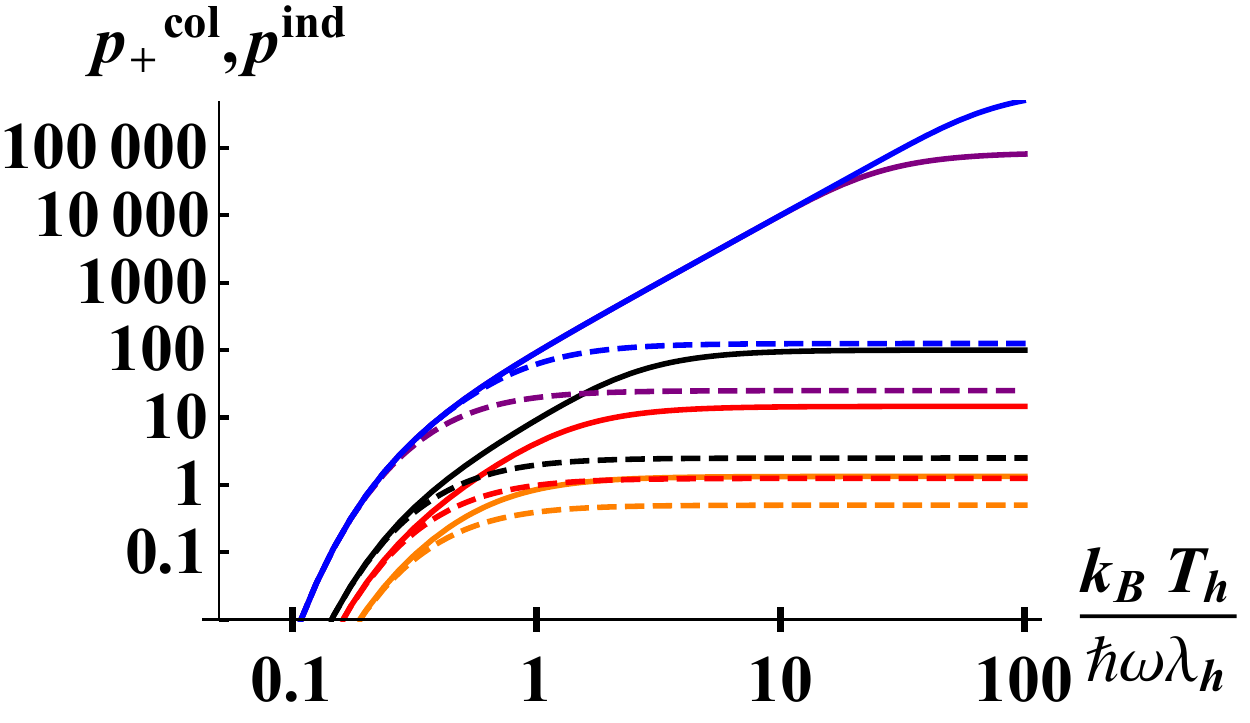}\\
(b)\includegraphics[width=8cm, height=5.5cm]{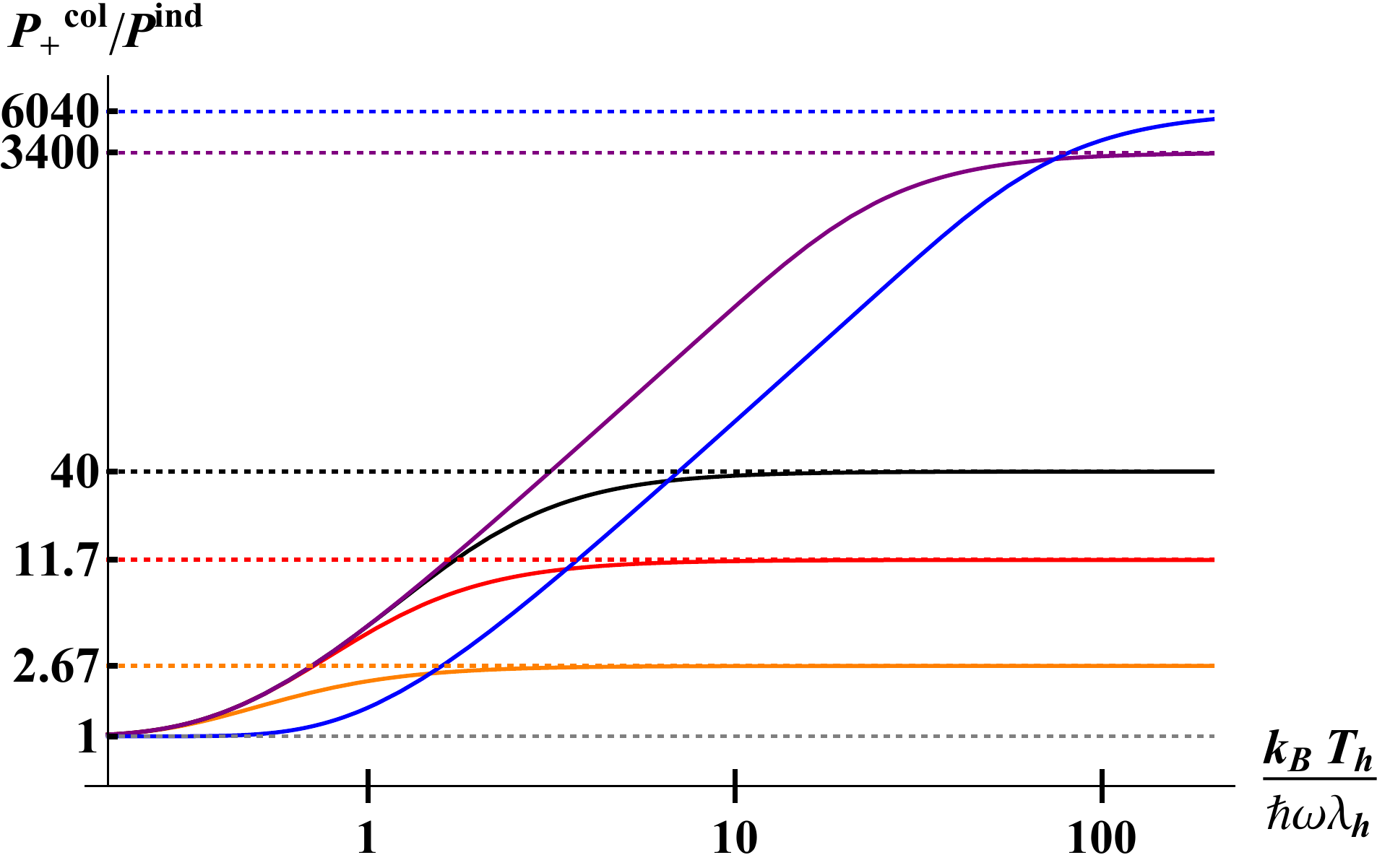}
\caption{(a) Plots of $p^{\rm col}_{+}:= {\cal P}^{\rm col}_{+}\tau_{\rm ind}/(\Delta \eta \lambda_h^2\Delta\beta)$ (solid curves) and $p^{\rm ind}:={\cal P}^{\rm ind}\tau_{\rm ind}/(\Delta \eta \lambda_h^2\Delta\beta)$ (dashed curves) as a function of $k_BT_h/\hbar\omega \lambda_h$ for fixed $\Delta \eta$, $\lambda_h$, $\Delta \beta = \beta_c-\beta_h$, and for ensembles containing $n=2$ (orange curve), $n=5$ (red curve), $n=10$ (black curve), $n=100$ (purple curve) spins $s=1/2$. The blue curves represent an ensemble of $n=100$ spins $s=3/2$. (b) Plots in log-log scale of the ratio ${\cal P}_{+}^{\rm col}/{\cal P}^{\rm ind}$ with the same color code as in the panel (a). The horizontal dotted lines indicate the asymptotic behaviours which follow the analytical value $n(ns + 1)/(s + 1)$.} 
\label{power}
\end{figure}

The length time of each cycle is mainly determined by the thermalisation time -- the isentropic strokes can be made in principle on a timescale much smaller than the thermalisation time as long as one considers a driving such that $[H(t),H(t')]=0$ for all $t, t'$, which we assumed here. The thermalisation time can be estimated from the dynamics and can be very different between collective and independent bath coupling. This phenomenon was indeed exploited in \cite{Kloc_2019}.
In Appendix \ref{coltimescale} we show that for spin ensembles initially in a thermal state at inverse temperature $\beta_0$ such that $\hbar\omega|\beta_0|\gg1$ the timescale to reach the steady state through collective interactions is at least $n$ times shorter than the equilibration timescale for independent dissipation. Although stemming from collectively-enhanced dissipation rates as superradiance does, this accelerated equilibration is a different phenomenon. In particular, it happens for any state in the symmetrical subspace ($J=ns$) whereas superradiant states are limited to values of $m$ close to zero. 
Thanks to this accelerated equilibration, the timescale $\tau_{\rm col}$ of the cycles of the collective spin engine can set to be $n$ times shorter than $\tau_{\rm ind}$, the timescale of the cycles of the independent spin engine.   
Then, with $\tau_{\rm col} = \tau_{\rm ind}/n$ the output power of the Otto machines are given by
 \be
{\cal P}^{\rm ind} = \frac{1}{\tau_{\rm ind}}\Delta \eta \lambda_h^2 (\beta_c-\beta_h)\frac{nC_s(\theta_h)}{k_B\theta_h^2} + {\cal O}(\Delta \eta^2)
 \ee
 and 
  \be
{\cal P}_{+}^{\rm col} = \frac{n}{\tau_{\rm ind}}\Delta \eta \lambda_h^2 (\beta_c-\beta_h)\frac{C_{ns}(\theta_h)}{k_B\theta_h^2} + {\cal O}(\Delta \eta^2).
 \ee
In particular we have 
\be
\frac{{\cal P}_{+}^{\rm col}}{{\cal P}^{\rm ind}}~~ \underset{T_h\gg \lambda_hT_{\rm cr}(n,s)}{\sim}~ n\frac{ns+1}{s+1}.
\ee
Note that due to the same issue of saturation commented above, collective effects still cannot be used to reach asymptotically the Carnot efficiency at finite output power. However, we have now that at fixed $T_h$, when the size of the working medium is much larger than the critical size $n_{\rm cr}(T_h,\lambda_h,s)$, the output powers of the two machines become equivalent (instead of $W^{\rm th}/W^{\rm col}_{+} \sim n$ for the extracted work per cycle). Therefore, in terms of output {\it power}, the collective Otto engine performs always better than or equal to the independent Otto engine. To illustrate this important point, the plots of $p_{+}^{\rm col} := {\cal P}_{+}^{\rm col}\tau_{\rm ind}/(\Delta \eta \lambda_h^2\Delta\beta)$, $p^{\rm ind}:= {\cal P}^{\rm ind}\tau_{\rm ind}/(\Delta \eta \lambda_h^2\Delta\beta)$ and ${\cal P}_{+}^{\rm col}/{\cal P}^{\rm ind}$ are shown in Fig. \ref{power} for several size of the spin ensemble.

Finally, it is worth mentioning that the $(n+2)/3$-fold power enhancement reported in continuous machines using collective coupling with spins $1/2$ \cite{Niedenzu_2018} do not coincide with the the above $n(ns+1)/(s+1)$-fold enhancement. However, it coincides with the first enhancement mentioned above for the extracted work per cycle. This emphasises that there are two different sources of enhancement in our model. One coming from alterations of the steady state properties of the spin ensemble due to collective coupling and bath-induced coherences \cite{bathinducedcoh,bathinducedcohTLS,negcontrib}, which appears in continuous engines \cite{Niedenzu_2018,Niedenzu_2015} in the form of a kind of ``steady-state'' superradiance \cite{Gross_1982}. The other source of enhancement comes from the fast equilibration mentioned above, and is not present in continuous machines.
Note that this manifest non-equivalence between cyclic and continuous engines is not in contradiction with \cite{Uzdin_2015} since the setup considered here and in \cite{Niedenzu_2015,Niedenzu_2018} do not belong to the small action regime used in \cite{Uzdin_2015} to derive the equivalence between cyclic and continuous engines.

\section{Conclusion and final remarks}\label{concl}
We show that collective coupling between spins and a thermal bath leads to a collective heat capacity which takes very different values when compared to the independent heat capacity -- when all spin interact independently with the thermal bath.
Beyond being a curiosity by itself, we show two applications.  The first one is related to quantum thermometry. We show that using a probe made of spins collectively coupled to the sample can bring large precision enhancements at high temperature, which can be of interest in some technological or bio-medicinal applications. In terms of spin number $n$ contained in the ensemble, the precision scales as $\Delta T \sim 1/n$, corresponding to the famous Heisenberg scaling. Interestingly, such Heisenberg scaling is achieved with a probe initially prepared in a pure classical state (namely thermal states at inverse temperature satisfying $\hbar\omega|\beta_0|\gg 1$) whereas it usually requires initial states with genuine quantum characteristics like squeezing \cite{CLL_2013, CLL_2016} or entanglement \cite{Escher_2011,Escher_2011b,Giovanetti_2006}. 
However, the price to pay is a potentially complex experimental setup in order to realise collective interactions between the spins and the sample, as commented in Section \ref{Experiment}.  
We also determine the critical temperature $T_{\rm cr}$ below which collective couplings stop being beneficial.  

In a second part of the paper, we apply the results on collective heat capacity to Otto engines using an ensemble of $n$ spins of dimension $s$ as working medium and operating close to the Carnot efficiency. 
Our results show that the output power of a collective spin engine (when the spins interact collectively with the baths) is always larger than or equal to the output power of an independent spin engine (when the spins interact independently with the baths). The enhancement is two-fold: a steady-state effect  relying on bath-induced coherences bringing a $(ns+1)/(s+1)$-fold enhancement and a fast equilibration bringing an additional $n$-fold enhancement.  
Combining both, the largest enhancements happen at high hot bath temperatures, reaching asymptotically levels of $n(ns+1)/(s+1)$-fold enhancements.

One can wonder if similar enhancements can happen in low-dissipative Carnot engines since the crucial role of heat capacity in such engines was recently pointed out \cite{Abiuso_2019}. While one can show that the role of the heat capacity remains prominent for an ensemble of spins interacting independently with the thermal baths, it is not obvious that it can be extended to spins interacting collective with the baths. Indeed, the dynamics of ${\rm Tr}\rho J_z$ does not follow a simple exponential decay so that a more complex treatment is needed \cite{Abiuso_2019}, involving numerical methods. Such complexity emphasises that although it is quite intuitive to see the heat capacity playing a central role in thermal machines, for certain designs it is a highly non-trivial conclusion. 
Still, it would be interesting to try to overcome these obstacles to investigate collective couplings in low-dissipative Carnot engines as new phenomena might emerge.


Finally, one could say that the performances of the best case scenario, when the initial state belongs to the symmetrical subspace, can be reproduced by a larger single spin of dimension $ns$. Furthermore, one could add that since the dissipative dynamics of a single spin does not involve generation of coherences or quantum correlations, there is nothing genuinely quantum in the performances of the spin ensemble collectively coupled to the sample or thermal bath. 
While it is true that the best-case scenario performances of the spin ensemble reproduce the ones of a single spin of dimension $ns$, the comparison is a bit unfair because the systems are not the same. Classical and quantum performances should be compared with the same systems and the same resources. Beyond that, in any experimental implementations the size of the number of spins are limited. If one wants to increase the performances beyond this classical limitation, one can use collective bath couplings. 
Overall, our results advocate for increasing efforts towards experimental realisations of collective couplings.

\acknowledgments 
CLL wants to thank Mohammad Mehboudi for interesting and enthusiastic discussions related to the thermometry part of the paper. 
This work is based upon research supported by the South African Research Chair Initiative, Grant No. UID 64812 of the Department of Science and Technology of the Republic of South Africa and National Research Foundation of the Republic of South Africa.

\appendix
\numberwithin{equation}{section}

\section{Maximal precision from collective steady states}\label{appcolqfi}
In this section we show that the maximal precision related to the temperature estimation when using collective interaction with the sample are determined by the collective heat capacity $C^{\rm col}(\beta)$. In order to show that we have to compute the quantum Fisher information associated with the collective steady state, $\rho^{\rm ss}(\beta)$. The first issue is that there is no general explicit expression of the quantum Fisher information for mixed states. 

Before continuing we must introduce some concepts of quantum metrology. The quantum Fisher information is defined as the maximum over all possible measurements -- described by a POVM -- of the Fisher information \cite{Fisher_1925}. The Fisher information represents the amount of information about the parameter of interest contained in the output statistics of a given measurement. For instance, let us consider a measurement described by the operators $\{E(m)\}_m$ forming a positive-operator-valued measure (POVM). The probability distribution of the result $m$ is given by
\be
p(m|T) := {\rm Tr} \rho^{\rm ss}(T)E(m).
\ee
The information about the temperature $T$ that one can extract from the measurement outputs statistics is the Fisher information
\be
{\cal F}_{E(m)}(T) = \sum_m \frac{1}{p(m|T)}\left(\frac{\partial p(m|T)}{\partial T}\right)^2.
\ee 
The quantum Fisher information can be formally defined as ${\cal F}^{\rm col}(T) := {\rm Max}_{\{E(m)\}}{\cal F}_{E(m)}(T)$. However, the direct maximisation over all possible measurements is usually not tractable. One alternative involves purifications in larger Hilbert spaces \cite{Bruno_2011, CLL_2013}. An other alternative is to use the ``symmetric logarithmic derivative'' operator of $\rho^{\rm ss}(\beta)$, implicitly defined by \cite{Braustein_1994, Helstrom_1976, Holevo_1982}
\be\label{sld}
\frac{\partial }{\partial T}\rho^{\rm ss}(\beta) = \frac{1}{2} L_T \rho^{\rm ss}(\beta) +\frac{1}{2}\rho^{\rm ss}(\beta)L_T,
\ee
which has the interesting property of providing the quantum Fisher information through the relation
\be\label{qfisld}
{\cal F}^{\rm col}(T) = {\rm Tr} \rho^{\rm ss}(\beta) L_T^2.
\ee
 For thermal states one can easily verify that $L_T = \frac{(\hbar\omega)^2}{(k_BT)^2}(J_z - \langle J_z \rangle)$, which leads immediately that the maximal precision is determined by the variance of the energy or equivalently by the heat capacity -- for thermal states. Beyond thermal states, it is in general very hard to find one symmetric logarithmic derivative operator, and unfortunately this includes the case of collective steady states. Still, one can show that
\be
\frac{\partial }{\partial T}\rho^{\rm ss}(\beta) = \sum_{J=J_0}^{ns} \sum_{i=1}^{l_J} \frac{p_{J,i}}{2} \big( L_{T,J} \rho_{J,i}^{\rm th}(\beta) + \rho_{J,i}^{\rm th}(\beta)L_{T,J}\big),
\ee
 with $L_{T,J} = k_B^{-1}T^{-2} [\hbar\omega J_z - e_J(\beta)]$, which is not of the form \eqref{sld} but instead a sum of symmetric logarithmic derivative operators acting on each eigenspace $J$. Thus, the relation \eqref{qfisld} does not hold automatically. Even though, using  twice the Cauchy-Schwarz inequality one can show 
 \bea\label{upperb}
 {\cal F}^{\rm col}(T) &\leq& \sum_{J=J_0}^{ns} \sum_{i=1}^{l_J} p_{J,i} {\rm Tr} \rho_{J,i}^{\rm th}(\beta) L_{\beta,J}^2\nn\\
 && = T^{-2} \left[ (\hbar\omega\beta)^2 \langle J_z^2\rangle_{\rho^{\rm ss}(\beta)} -\beta^2 \overline{e^2_J(\beta)}\right]\nn\\
 && \leq T^{-2} (\hbar\omega\beta)^2 \Delta^2 J_z
 \eea
 where $\overline{e^2_J(\beta)} := \sum_{J=J_0}^{ns}\sum_{i=1}^{l_J} p_{J,i}e_J^2(\beta)$ and $\Delta^2 J_z :=  \left[ \langle J_z^2\rangle_{\rho^{\rm ss}(\beta)}- \langle J_z\rangle_{\rho^{\rm ss}(\beta)}^2\right]$.
 
 Now that we have an upper bound, the next step is to show that there exists one measurement $\{E(x)\}_x$ such that the associated Fisher information ${\cal F}_{\{E(x)\}}(T)$ saturates the upper bound. Naturally, one can think of energy measurements since it is the best measurement for thermal states \cite{Mehboudi_2019}. The energy measurements is described by the following POVM $\Pi_{m} := \sum_{J=|m|}^{ns}\sum_{i=1}^{l_J} |J,m\ket_i\bra J,m|$, with $m \in [-ns;ns]$ denoting the eigenvalues of the Hamiltonian $H_S = \hbar\omega J_z$ of the spin ensemble. One can show that ${\cal F}_{\Pi_m}(t)$ does not reach the upper bound \eqref{upperb}. 
 
 There is indeed one measurement which can extract more information about $T$ than the energy measurement. This is the measurement described by $\Pi_{J,m,i} :=|J,m\ket_i\bra J,m|$, which corresponds to the projection onto the collective states $|J,m\ket_i$. One can show easily that 
 \be
 {\cal F}_{\Pi_{J,m,i}} =   T^{-2} \left[ (\hbar\omega\beta)^2 \langle J_z^2\rangle_{\rho^{\rm ss}(\beta)} - \beta^2\overline{e^2_J(\beta)}\right].
 \ee
This allows us to conclude that the upper bound \eqref{upperb} is indeed an equality,
 \be
 {\cal F}^{\rm col}(T)  = T^{-2} \left[ (\hbar\omega\beta)^2 \langle J_z^2\rangle_{\rho^{\rm ss}(\beta)} -\beta^2 \overline{e^2_J(\beta)}\right].
 \ee
Finally, one can also show that the collective heat capacity $C^{\rm col}(\beta)$, defined in \eqref{cindist}, can alternatively be expressed as $C^{\rm col}(\beta) = k_B \left[(\hbar\omega\beta)^2 \langle J_z^2\rangle_{\rho^{\rm ss}(\beta)} - \beta^2\overline{e^2_J(\beta)}\right]$, so that
\be
 {\cal F}^{\rm col}(T)  = k_B\beta^2 C^{\rm col}(\beta),
 \ee
as announced in the main text. 

One remark is in order. The optimal measurement yielding an information equal to the quantum Fisher information is $\{\Pi_{J,m,i}\}_{J,m,i}$, which is a non-local measurements. Therefore, it is not really realistic to consider that it is actually possible to experimentally saturates the quantum Fisher information and the estimate the temperature's sample at the corresponding precision. However, in the best case scenario where the spin ensemble initially belongs to the symmetrical subspace, like in particular for thermal states at extreme inverse temperature $|\beta_0| \gg 1/\hbar\omega$, the energy measurement $\{\Pi_n\}_m$ indeed yields an information equal to the quantum Fisher information. Then, the precision announced in the main text Eq. \eqref{maxprecision} is achievable experimentally, at least from the point of view of the measurements.

\section{Collective dissipation timescale}\label{coltimescale}
In this section we show that, for initial state belonging to the symmetrical subspace, collective interaction with the bath yields a dissipation timescale $n$ times shorter than independent dissipation. 
We start from the dynamics of the collective dissipation provided in Eq. \eqref{colme}. 
 Assuming that the ensemble is initially in a thermal state, it is initially diagonal and will remain diagonal in the collective basis $\{|J,m\ket_i\}$, $J\in[J_0;ns]$, $m \in [-J;J]$, $i\in [1;l_J]$. Therefore, the dynamics is given by the populations $ p_{J,m,i} :=~_i\bra J,m|\rho|J,m\ket_i $ only. Using the relation \cite{Sakurai_Book} 
 \be
J^{\pm}|J,m\ket_i =\hbar \sqrt{(J\mp m)(J\pm m+1)} |J,m\pm 1\ket_i,
\ee
one obtains
\bea\label{colpop}
\dot p_{J,m,i} & =& G(\omega)[(J-m)(J+m+1)p_{J,m+1,i}\nn\\
 && \hspace{1cm}- (J+m)(J-m+1)p_{J,m,i}] \nn\\
 &+& G(-\omega)[(J+m)(J-m+1)p_{J,m-1,i} \nn\\
 &&\hspace{1cm}- (J-m)(J+m+1)p_{J,m,i}],
\eea
where $G(\omega):=\Gamma(\omega) + \Gamma^{*}(\omega)$, and $\Gamma(\omega)$ is the ``half Fourier transform'' of the bath correlation function introduced in \eqref{colme}.
By contrast, the dynamics of the independent dissipation follows 
\bea\label{indepme}
\frac{ d \rho}{dt} &=& \Gamma(\omega) \sum_{i}^n\left(j^{-}_i\rho j^{+}_{i} - j_i^{+} j_i^- \rho\right) \nn\\
&&+ \Gamma(-\omega) \sum_{i}^n\left(j_i^{+}\rho j_i^{-} - j_i^{-} j_i^{+} \rho\right)  + {\rm h.c.}.
\eea
Again, since the ensemble is assumed to be initially in a thermal state, all coherences are and remain null (this would not be true for collective dissipation, reason why the collective basis was used there). Then, the independent dissipation is described by the populations only,
\bea\label{indpop}
&&\dot p_{m_1,...,m_n} :=\bra m_1,...,m_n|\dot \rho|m_1,...,m_n\ket \nn\\
&& = G(\omega)\sum_{i=1}^n[(s-m_i)(s+m_i+1)p_{m_1,...,m_i+1,...,m_n}\nn\\
 && \hspace{1.5cm}- (s+m_i)(s-m_i+1)p_{m_1,...,m_i,...,m_n}] \nn\\
 &&+ G(-\omega)\sum_{i=1}^n[(s+m_i)(s-m_i+1)p_{m_1,...,m_i-1,...,m_n} \nn\\
 &&\hspace{1.5cm}- (s-m_i)(s+m_i+1)p_{m_1,...,m_i,...,m_n}].
\eea
One can see that the non-zero coefficients appearing in \eqref{colpop}, which determines the rate of each transition and consequently the timescale of the dissipation, range from $2G(\pm \omega)J$ to $G(\pm \omega)J(J+1)$ for $m \in [-J,J]$. By contrast, the rates of transition in \eqref{indpop} range from $2G(\pm\omega)s$ to $G(\pm\omega)s(s+1)$. One recovers in particular that the equilibration timescale (for independent dissipation) is of the order $G(\omega)^{-1}\sim (g^2\tau_c)^{-1}$, where $\tau_c$ is the bath correlation time. Since $J$ can take value  from $0$ or $1/2$ to $ns$, the timescale to reach the steady state is in general of the same order or even larger for collective dissipation than for independent dissipation. However, for a thermal state at initial inverse temperature $\hbar\omega|\beta_0|\gg 1$ (or more generally belonging to the symmetrical subspace), all components of $J<ns$ are null implying that all transition rates involved in the collective dissipation are at least $n$ times larger than the transition rates of the independent dissipation. Then, for initial states such that  $\hbar\omega|\beta_0|\gg 1$ the collective dissipation happens on a timescale at least $n$ times faster than the independent dissipation. Note that this interesting result shares some similarities with superradiance \cite{Gross_1982}. Indeed, both phenomena rely on the collective damping rates appearing in \eqref{colpop}, larger than the independent damping rates of \eqref{indpop} for states confined to the symmetrical subspace ($J=ns$). However, superradiance is an increase of the rate of energy change whereas this ``super-equilibration'' is purely about the dynamics of the spin ensemble's state itself. Note also that superradiance appears for states $|J=ns,m\ket_i$ with $m$ close the zero while the accelerated equilibration occurs for any states in the symmetrical subspace.  
As a consequence of this accelerated equilibration, the timescale $\tau_{\rm col}$ of one cycle of the collective Otto machine can be reduced by a factor $n$ compared to the timescale $\tau_{\rm ind}$ of the independent Otto machine.

\section{Note on the stability of the collective steady state}\label{note}
For very small imperfections like tiny spin-spin interactions or inhomogeneities and disorder altering the local energy levels of each spin, both tending to break down the spin exchange symmetry (or equivalently, the spin indistinguishability), it was shown in \cite{bathinducedcoh} that the steady state \eqref{colss} was still reached as long as $\delta \ll g^2\tau_c$, where $\delta$ stands for the order of magnitude of the energy involved in the imperfections, $g$ is the coupling strength with the bath, and $\tau_c$ is the bath correlation time. Furthermore, in the applications to thermometry and engines we are mostly interested in initial states such that $p_{J=ns}\simeq 1$ for which the equilibration time is of the order of $(n g^2\tau_c)^{-1}$ (see appendix \ref{coltimescale}). Therefore, for such initial states, the condition on the magnitude of the imperfections is relaxed to $\delta \ll n g^2\tau_c$.

\section{Work per cycle for Otto engine operating near the Carnot bound}\label{wpercycle}
In this section we detail briefly the derivation of the expression of the output work per cycle. During the isochoric stroke in contact with the hot bath, the spin ensemble is brought to the state $\rho_1:=\rho^{\rm ss}(T_h, \lambda_h)$. The next stroke is isentropic, preserving the state of the spin ensemble while realising the relaxation $\lambda_h \rightarrow \lambda_c$. Then, follows the second isochoric stroke, taking the spin ensemble to $\rho_2:=\rho^{\rm ss}(T_c, \lambda_c)$. The last isentropic stroke is a compression $\lambda_c \rightarrow \lambda_h$, closing the cycle. Note that, as mentioned in the main text, the initial weights $p_{J,i}$ are preserved throughout the cycles and have a crucial impact on the properties and performances of the engine.
The work $W^{\rm col}$ extracted per cycle by the engine is the sum of the work realised during the two isentropic strokes,
 \bea
W^{\rm col} &=& {\rm Tr} \rho_1[H(\lambda_c)-H(\lambda_h)] + {\rm Tr} \rho_2[H(\lambda_h)-H(\lambda_c)] \nn\\
&=& \hbar\omega(\lambda_c-\lambda_h){\rm Tr} J_z[\rho^{\rm ss}(T_h,\lambda_h)-\rho^{\rm ss}(T_c,\lambda_c)]\nn\\
&=&(\lambda_c-\lambda_h)[E^{\rm ss}(\theta_h)-E^{\rm ss}(\theta_c)],
\eea 
where $\theta_x:= \lambda_x\beta_x$, for $x=h,c$. Note that the only way of having work extraction ($W^{\rm col}<0$) is with compression factors satisfying the condition $1< \frac{\lambda_h}{\lambda_c}<\frac{\beta_c}{\beta_h}$.
The work extraction efficiency is defined as $\eta:=-\frac{W^{\rm col}}{Q_h}$, where $Q_h= {\rm Tr} H(\lambda_h)[\rho^{\rm ss}(T_h, \lambda_h)-\rho^{\rm ss}(T_c, \lambda_c)]$ is the heat transferred from the hot bath to the spin ensemble. One recovers the usual expression for the efficiency, $\eta= 1- \frac{\lambda_c}{\lambda_h}$, and the difference with the Carnot efficiency $\eta_c=1-\frac{\beta_h}{\beta_c}$ is $\Delta \eta = \frac{\lambda_c}{\lambda_h} -\frac{\beta_h}{\beta_c}$. One can rewrite the output work in terms of $\Delta \eta$ to obtain
\be
W^{\rm col} = \lambda_h (\Delta\eta +\frac{\beta_h}{\beta_c} - 1)[E^{\rm ss}(\theta_h)-E^{\rm ss}(\theta_h + \lambda_h\beta_c\Delta\eta)].
\ee
Taking the limit of near Carnot efficiency, $\Delta \eta\rightarrow 0$, the output work takes the form
\be
W^{\rm col} = -\Delta \eta \lambda_h^2(\beta_c-\beta_h)\frac{C^{\rm col}(\theta_h)}{k_B\theta_h^2} +{\cal O}(\Delta\eta^2).
\ee

One can repeat the same reasoning with the alternative situation where each spin interacts independently with the baths. One obtains the expression found in \cite{Campisi_2016}
\be
W^{\rm ind} = -\Delta \eta \lambda_h^2(\beta_c-\beta_h)\frac{C^{\rm ind}(\theta_h)}{k_B\theta_h^2} +{\cal O}(\Delta\eta^2).
\ee
These are the expressions used in the main text. Note that in the main text we consider implicitly the absolute value of the extracted work so that the front minus sign is dropped in the above expressions.

\end{document}